# Single ion implantation for single donor devices using Geiger mode detectors

E. Bielejec, J. A. Seamons[+] and M. S. Carroll

Sandia National Laboratories, P.O. Box 5800, Albuquerque, NM 87185-1056, USA

E-mail: esbiele@sandia.gov

**Abstract.**
Electronic devices that are designed to use the properties of single atoms such as donors or defects have become a reality with recent demonstrations of donor spectroscopy, single photon emission sources, and magnetic imaging using defect centers in diamond. Ion implantation, an industry standard for atom placement in materials, requires augmentation for single ion capability including a method to detect a single ion arrival. Integrating single ion detection techniques with the single donor device construction region allows single ion arrival to be assured. Improving detector sensitivity is linked to improving control over the straggle of the ion as well as providing more flexibility in lay-out integration with the active region of the single donor device construction zone by allowing ion sensing at potentially greater distances. Using a remotely located passively gated single ion Geiger mode avalanche diode (SIGMA) detector we have demonstrated 100% detection efficiency at a distance of >75 μm from the center of the collecting junction. This detection efficiency is achieved with sensitivity to ~600 or fewer electron-hole pairs produced by the implanted ion. Ion detectors with this sensitivity and integrated with a thin dielectric, for example 5 nm gate oxide, using low energy Sb implantation would have an end of range straggle of <2.5 nm.

Significant reduction in false count probability is, furthermore, achieved by modifying the ion beam set-up to allow for cryogenic operation of the SIGMA detector. Using a detection window of 230 ns at 1 Hz, the probability of a false count was measured as $\sim 10^{-1}$ and $10^{-4}$ for operation temperatures of ~300K and ~77K, respectively. Low temperature operation and reduced false, "dark", counts are critical to achieving high confidence in single ion arrival. For the device performance in this work, the confidence is calculated as a probability of >98% for counting one and only one ion for a false count probability of $10^{-4}$ at an average ion number per gated window of 0.015.

## 1. Introduction

Recently significant progress has been demonstrated in experimentally realizing electronic or optoelectronic device structures using single donors or defect centers [1-9]. Ion implantation is an industry standard for placing atoms in the near surface region with high precision, for example, for controlled source/drain junctions in silicon complementary metal oxide semiconductor (CMOS) applications. Accurate and precise placement of exactly one ion using ion implantation represents a significant challenge because it requires verification of a single ion arrival, while also limiting straggle [6-7]. Straggle is a measure of the random displacement of the ion as it is stopped by the substrate and the straggle decreases with decreasing ion energy.

---

[+] Current Address: Space Dynamics Laboratory, North Logan Utah

One approach to ion detection is through sensing electron-hole (e-h) pair generation with a built-in detector near the single donor device construction region. The state-of-the-art for low energy single ion detection using a built-in p-i-n diode is signal-to-noise limited to an energy of ~10 keV P ions producing ~$10^3$ e-h pairs per implanted ion [7]. Batra *et al.*, [10] and Shinada *et al.*, [11] have alternatively demonstrated successful single ion detection by monitoring changes in transistor current after implanting Si, P, Xe, or Sb donors, 30 – 70 keV energies, into the channel of micron scale Si transistors. Increased sensitivity to lower energy ions is desirable to decrease straggle below the range of ~10 nm [6, 7] for single atom devices such as solid-state quantum bits [12].

Avalanche photodiodes (APD) have been used to sense single e-h pairs generated by the arrival of single photons through use of passively gated Geiger mode (GM) operation of the APD. This method was demonstrated for single ion detection at 300K [13]. Single ion Geiger mode avalanche diode (SIGMA) detectors operate by over-biasing past reverse breakdown for a short time period (which we call the gating window) during which the detector can be sensitive to a single electron or hole injected into the junction. The high fields produce an avalanche cascade leading to a Geiger signal. The gating window can coincide with a pulsed ion beam configuration to insure that the detector is on when an ion is expected to arrive. A challenge in this configuration is the high number of dark counts, miscounts due to thermally generated e-h pairs, that obscures the detection of a single ion arrival. Furthermore, extension of the lateral sensitivity using this approach, to-date, has not been well established. Increased flexibility in remote lateral sensing is important for future integration with single donor device construction regions.

We report a mapping of the remote lateral sensing of the SIGMA detector combined with demonstrating a significant reduction in dark count probability. The dark count reduction by three orders of magnitude was accomplished through the introduction of a 77K cooled stage. The mapping of the lateral sensitivity was determined using the ion beam induced current (IBIC) response of the detectors. The lateral sensitivity of the SIGMA detector at ~150 µm diameter has an estimated upper bound of 600 e-h pairs or less while maintaining 100% detection efficiency within the uncertainty of the measurement. The SIGMA sensitivity provides a path towards detecting lower energy ions (i.e., smaller straggle) and greater flexibility in the lay-out of single ion detector integration with the single donor device construction zones. The probability of counting one and only one ion is calculated to elucidate the dependence on the false count probability and average number of ions per gate window. An optimum of 0.015 ions per gate window is calculated for the measured $10^{-4}$ false count probability of the devices discussed here. At this optimum ion count per gate, a single ion can be implanted with a confidence of greater than 98%. Approaches to improve this confidence are also discussed.

## 2. Experiment

The SIGMA detectors were probed with an $H^+$ beam at energies from 75 to 250 keV. A focused beam was used for the experiments and was produced using a modified High Voltage Engineering Europa implanter. The beam-line end-station was, furthermore, equipped with a temperature controlled sample stage to allow operation of the SIGMA detector at either room (300K) or liquid nitrogen (~77K) temperatures. The cooling was provided by flowing liquid nitrogen through a cold finger on which the sample was placed.

The microbeam was run in two modes of operation: broad pulsed beam and focused pulsed microbeam. For the experiments described in the first part of this paper we have used a broad beam setup with a ~50 µm beam spot with an average of 1 ion/pulse for a pulse length of 1 µs, this corresponds to an on target beam current of ~0.15 pA. The incident ions obey a Poisson distribution resulting in a 63.4% probability that one or more ions will arrive during the gate time. The focused microbeam mode, spot size of ~800 x 800 $nm^2$, allows spatial mapping of the Geiger mode detection efficiency across the detector and its lateral extent using IBIC as well as Geiger mode collection to characterize lateral remote ion detection. A rate of ~100-10,000 ions/s was used in the focused microbeam configuration.

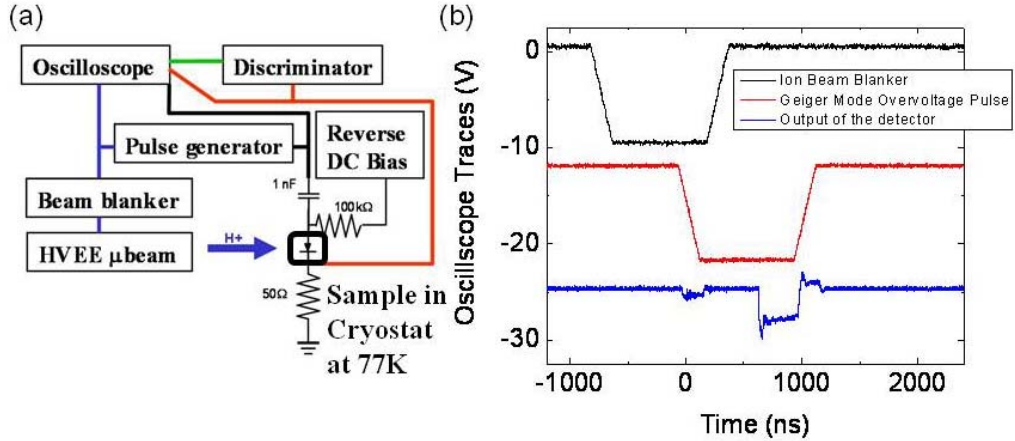

**Figure 1.** (a) The circuit including passive quenching used in these measurements. (b) The oscilloscope traces showing the ion gate pulse (black), the SIGMA gate over pulse times (red, offset by 12 V), the Geiger avalanche pulse on the output from the SIGMA detector (blue, scaled by 250 and offset by 25 V) each has been scaled and off-set for clarity. (Color online only)

The SIGMA detectors used in these experiments were fabricated using process steps from a 0.35 μm CMOS fabrication line at Sandia National Laboratories (SNL). More details about the diode fabrication can be located in reference 13. We characterized the SIGMA detector both below breakdown and in GM operation. Dark and photocurrent versus voltage curves were used as a diagnostic of the APD. The GM operation consists of reverse biasing the device near breakdown and then pulsing beyond breakdown for short periods of time (230 ns or 1 μs) called detection windows. During this over-voltage pulse the avalanche breakdown is sensitive to and can be triggered by one or more free carriers in active region [14]. False counts are generated predominantly by free carriers diffusing into or tunneling through the junction that are not from the ion stopping process (e.g., thermal generation and trap assisted tunneling). The avalanche breakdown in the junction is subsequently quenched after the over-voltage is switched off. The circuit used in this measurement, Fig. 1a, and a typical measurement trace on the oscilloscope, Fig. 1b, show the ion beam blanker pulse, the SIGMA over-voltage pulse, and the SIGMA Geiger avalanche output pulse from top to bottom, respectively.

For typical GM operation the reverse bias was set to 145 V (below breakdown) and an additional 10 V over-voltage pulse, for 1 μs with rise and fall times of 120 ns, was applied to the device at a frequency of 10 Hz. The output from the SIGMA detector is then sent to a discriminator, which is set to detect the Geiger avalanche pulse, resulting in a digital output. For this experiment the ions are incident outside the active region of the SIGMA detector. The electrons created by the ion beam induced ionization diffuse into the active region and trigger the Geiger avalanche pulse. There is, therefore, a delay after the ion gating and before the SIGMA detector over-voltage pulse to compensate for the time of flight of the ions and to allow the carriers created in the Si to diffuse into the active region as indicated in Fig. 1b.

## 3. Results and discussion

Dark counts are shown as a function of detection window frequency and temperature in Fig. 2. In this case, the over-voltage pulse of 6.5 V past breakdown was applied for 230 ns and the measurement was repeated $10^4$ times. The shortest gate time achieved 230 ns, was limited by rise/fall

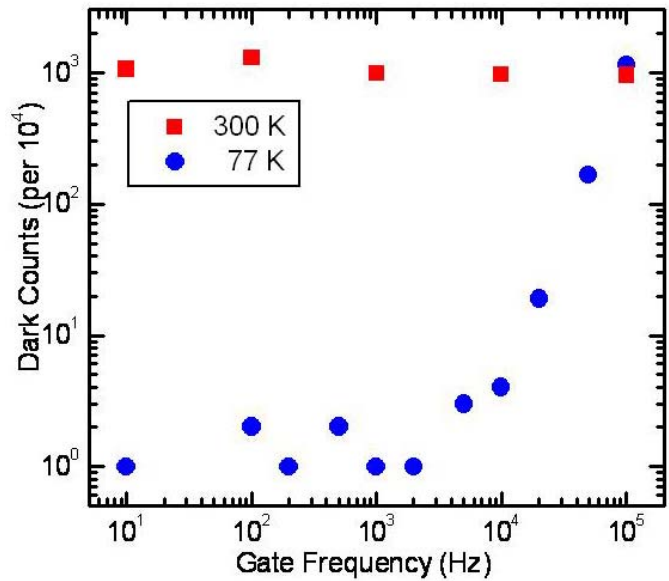

**Figure 2.** The dark counts from $10^4$ gatings of the SIGMA detector with individual durations of 230 ns, pulsing 6.5 V past breakdown, as a function of the gate frequency taken at 300 K (squares, red) and $LN_2$ temperature (circles, blue). (Color online only)

times of the pulse generator and was chosen because to minimize false counts at 300K, which was otherwise saturated at $10^4$ when a gate time of 1μs was used. This figure demonstrates that for gate frequencies less than ~2 kHz the dark counts are reduced by three orders of magnitude by lowering the temperature to ~77K.

At gate frequencies higher than 2 kHz it is presumed that trapped charge after an avalanche does not have sufficient time to discharge prior to the next over-voltage pulse, therefore the number of dark counts increases dramatically. This is a well know phenomena called afterpulsing [14, 15], which leads to complex optimization trade-offs between window detection time, pulse frequency, and device design. To achieve a high confidence that only one ion arrives, the average flux can be reduced to suppress the probability of two or more arrivals relative to a single ion arrival. As the flux is decreased, more implant windows will have no ions in the gate time producing an increase in necessary attempts at a single location before a single ion is implanted and detected. Therefore, increased detector frequency is desirable to maintain a relatively short dwell time at any given implant location while achieving a high confidence in the arrival of a single ion.

In Fig. 3a we plot the single ion detection efficiency at ~77K as a function of the delay time between the ion gate and the SIGMA over-voltage pulse for 250 keV $H^+$ ions implanted outside the active region of the junction. The approximate location of the ion beam spot is shown in the Fig. 3c. The detection efficiency is the probability of a count per ion arrival. It is calculated by dividing the sum of measured counts per second by the average number of ions implanted within that second. The average number of ions per second is the operating implant gate frequency multiplied by the nonzero ion arrival probability. The non-zero ion arrival probability is calculated assuming a Poisson distributed ion flux. The delay time necessary to achieve peak detection efficiency is a combination of the time-of-flight of the ions and the electronics response time, the long tail corresponds to a measure of the minority carrier lifetime, see reference 13 for more details.

The peak detection efficiency is also shown as a function of incident energy of the $H^+$ ion implanted in Fig. 3b. The detection efficiency is 100% within the uncertainty of the measurement for the ion energies greater than ~125 keV. The uncertainty in the confidence intervals shown in Fig. 3a and 3b results from a combination of the false count probability and the uncertainty in number of ions per window. Tighter confidence intervals are frustrated by degradation of the detector after a small number of ion strikes (on order of 1000), which limits the statistical sample that can be taken on one detector.

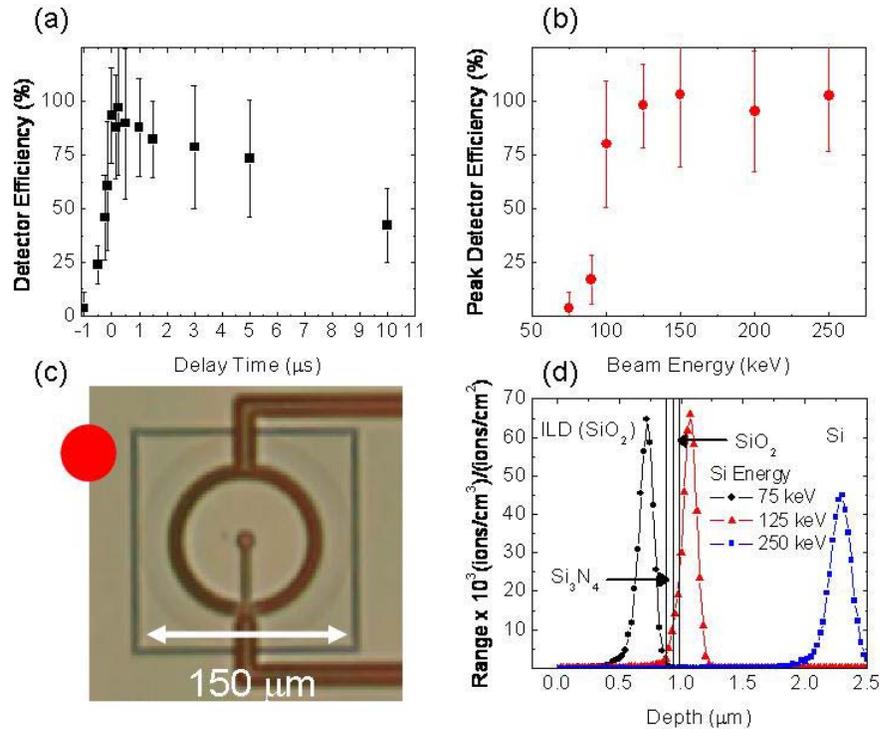

**Figure 3.** (a) The 250 keV H$^+$ ion detection efficiency of the SIGMA detector plotted as a function of the delay time between the ion gate and the GM over-voltage pulse. (b) The peak H$^+$ ion detection efficiency of the SIGMA detector plotted as a function of the implant energy. (c) A micrograph of the SIGMA detector with the location of the single ion implants within the red circle. (d) A SRIM plot indicating that the lowest beam energy necessary to create e-h pairs within the active region of the device is ~125 keV. (Color online only)

The decrease in the detector efficiency as the ion energy is lowered is due to the deceasing range of the incident ions, which leads to fewer e-h pair generation in the Si that can diffuse to the SIGMA detector junction. Calculations with SRIM show that H$^+$ ions will range out in the dielectric over-layers before reaching the Si bulk for energies less than 125 keV [16], see Fig. 3d. From SRIM calculations for 125 keV H$^+$ we find that approximately 1630 +/- 120 e-h pairs are produced in the Si body.

A lower bound on the collection efficiency can be estimated using ion beam induced charge collection (IBIC) with zero bias at ~77K, Fig. 4a. The zero bias IBIC collection efficiency provides an order of magnitude estimate for the SIGMA collection efficiency at high bias near breakdown, if we assume that all carriers that reach the edge of the SIGMA depletion region are collected, and the diffusion length of the remotely generated electrons is much larger than the depletion region difference between low and high bias operation (i.e., the lateral extent of the depletion region). The depletion width is estimated to be ~5 μm assuming a 1-sided junction and the nominal background doping of 10$^{16}$ cm$^{-3}$ boron, The electron diffusion length is estimated to be 70 μm, much larger than the depletion width. The electron diffusion length was calculated assuming a diffusivity of 25 cm$^2$/s, based on the background doping of the substrate and a minority carrier lifetime of 1.9 μs measured previously in devices that only differ in location on the wafer [13]. Accurate determination of collection efficiency with higher bias IBIC is obfuscated by non-linear internal gain due to below high field impact ionization.

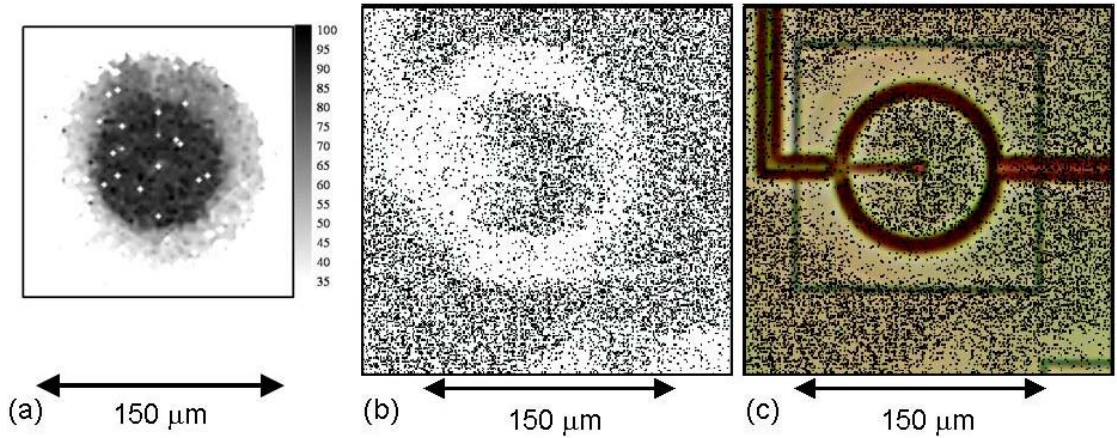

**Figure 4.** (a) The charge collection efficiency percentage of the APD in the SIGMA detector at zero bias as determined from IBIC with 250 keV $H^+$. Only inside the APD is the collection efficiency greater than the noise floor of 35%. (b) The digital IBIC map from the GM operation of the SIGMA at liquid nitrogen temperature. (c) GM-IBIC map superimposed upon a micrograph of the SIGMA detector. The low detection efficiency ring correlates with the anode metal contact and a heavily doped $n^+$ gettering ring. (Color online only)

Features of the device can readily be distinguished in the IBIC image. The visible circle of the detection efficiency is the inside of the outer metal ring of the SIGMA detector, Fig. 4a. In the interior of the SIGMA detector all the charge created is collected resulting in a collection efficiency of 100%. For 250 keV $H^+$, we use SRIM to predict that ~40,000 e-h pairs are produced in the Si body for a single ion implant. Collection efficiency is determined from knowing the number of electrons (and holes) generated (experimentally known from the ion rate) and measuring the number of charges collected (experimentally known from the IBIC measurement). The background signal at the edge of the IBIC sensitivity corresponds to ~14,000 e-h pairs. This result indicates that no more than ~35% of the generated carriers reach the detector junction when the ions strike in this remote location and that the IBIC's noise floor is ~14,000 e-h pairs.

In contrast, remote sensing and spatial mapping can also be done using Geiger mode operation. The SIGMA detector was found to be 100% sensitive from 250 keV to 125 keV for detection outside the active region at a diameter of approximately 150 μm (see Fig. 3b and 3c). The zero bias IBIC collection efficiency indicated that ~35% of the initial ion generated charge distribution reaches the detector junction (i.e., ~14,000 e-h pairs of the ~40,000 generated by the 250 keV case). Decreasing the energy to 125 keV, assuming a similar collection efficiency of 35%, corresponds to no more than ~600 e-h pairs generated at ~150 μm as the 125 keV $H^+$ ion strike produces only ~1,800 e-h pairs. This sensitivity to ~600 e-h pairs is over an order of magnitude smaller than the zero bias IBIC noise floor. We note that it is reasonable to expect that the detector efficiency will remain 100% down to much fewer e-h pair generation, as has been achieved for single photon detection using GM APDs; however, the thick dielectric over-layers in this device make it difficult to experimentally determine lower e-h pair generation ranges.

The spatial sensitivity of the Geiger mode approach was, furthermore, examined through imaging in an analogous way as IBIC. A combination of the IBIC raster scan with a pulsed focused ion beam and GM operation of the SIGMA detector. Fig. 4b, shows the GM ion beam induced current (GM-IBIC) of the SIGMA detector at liquid nitrogen temperature. As above we are gating both the 250 keV $H^+$ ion beam and the over-voltage pulse for 1 μs with a delay time of 500 ns. The dark count

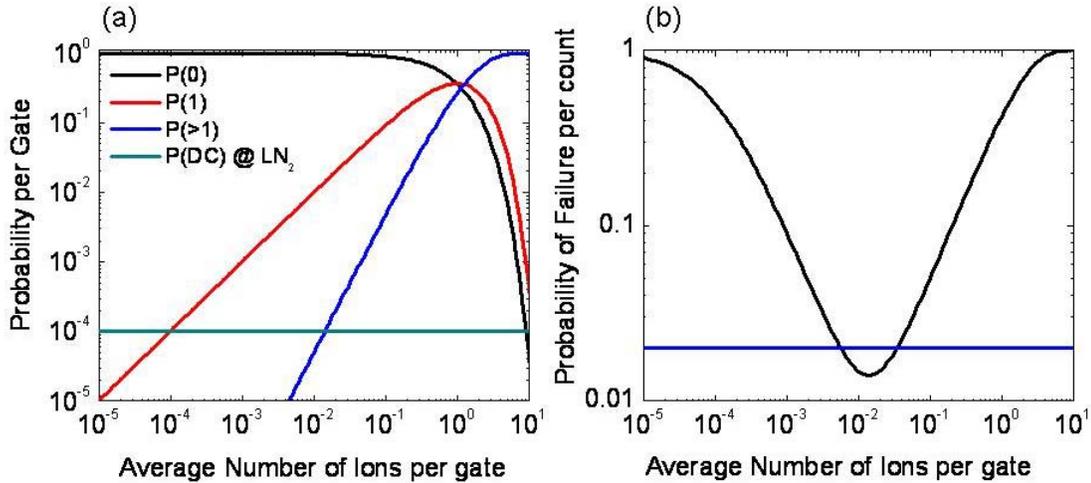

**Figure 5.** (a) Probability of 0 (black), 1 (red) and >1 (blue) ions as a function of the average number of ions per gate calculated using Poisson distributions. The probability of dark counts (DC) at ~77K is also plotted (green). (b) The probability per count that the detected signal on the SIGMA detector is not from one and only one ion is shown (solid black) as a function of the average number of ions per gate. We plot the probability of failure (1-probability of success) to clearly identify the desired operating regime. The blue line is a fixed probability of 2% failure for scale. (Color online only)

probability per gating is less than 1%. A micrograph of the actual SIGMA detector has been superimposed upon the GM-IBIC map in Fig. 4c. The detection is sensitive well outside of the active region of the SIGMA detector extending to a diameter larger than 150 μm. The low detection regions (e.g., ring and line) in the GM-IBIC correspond to metal contacts and a highly doped, deep, $N^+$ region in which the ions range out and therefore do not contribute to the collected charge.

The decrease in dark count rates at 77K and increase in sensitivity by using the SIGMA approach provide potential advantages both to assurance of single ion arrival (discussed below) and improved remote sensing (i.e., lay-out discussed above) as well as, in principle, offering improved precision in ion placement. The precision of the donor placement can be broken down into two different components; a lateral variation, primarily due to the spatial resolution of the defined aperture, for example see [6], and a longitudinal variation due to straggle of the implanted ion as it moves through the target material. The former is limited to ~10 nm using state-of-the-art electron beam lithography (EBL), whereas the latter is related to the energy of the ion beam and the material thickness, the composition of the target. The end position of the implanted atom is furthermore sensitive to the subsequent activation anneal condition. By using a SIGMA detector, the 100% detection efficiency to ~600 e-h pairs allows for lower energy implants than previously reported. In general the ability to reduce the number of e-h pairs needed to generate 100% detection efficiency will allow for tighter longitudinal precision (implant depth) in atom placement. We note, however, that randomness in the dopants position due to the diffusion length can become the dominant loss of longitudinal precision at these lower energies, although it is not clear what the relative contributions will be for an optimized thermal budget. The diffusion lengths of dopants are relatively well understood in the bulk, but critical diffusion parameters are not well established for dopants near the surface after implant for which the magnitude of transient enhanced diffusion and surface segregation [17] are not known. Therefore, at this time, although the increased sensitivity of the SIGMA detector approach provides the ability to remotely sense the ion arrival at greater distances than the present IBIC set-up, it is not clear that it provides improved ion placement precision due to other currently unknown limits from such as diffusion and EBL.

High percentage accurate confirmation of single ion arrivals, yield, is a critical parameter to optimize. The yield of single donor devices is linked to the interplay between detector efficiency, dark count probability and the probability of getting one and only one ion per detection window. The SIGMA detector cannot distinguish between false counts, single ion events or multiple ion events because of the detector's digital response. In Fig. 5a (the solid red line), the Poisson probability of one and no more than one ion, in a detection window is calculated as a function of the average number of ions per gate. The probability of zero ions (the solid black line) and greater than 1 ion implanted (the solid blue line) are shown to indicate that as we decrease the average number of ions per gate below 1 the probability of greater then 1 ions drops much faster than the probability of only 1 ion per gate. This indicates that increased confidence that one and only one ion arrives in a detection window can be increased significantly by decreasing the average number of ions per gate.

The average number of ions per gate can be decreased by either decreasing the flux of ions with a fixed gate time or by decreasing the gate time with a fixed flux or a combination of the two. Decreasing flux increases the probability of a dark count because on average more gating windows, or total time, is necessary to implant the ion. In contrast, if the gate time is shortened to reduce the average number of ions per gate (with fixed ion flux) then the probability of a dark count per gate decreases and balances out the extra number of gates necessary to implant one ion, Fig. 5a (green solid line for $LN_2$).

A characteristic time for a false or dark count, $\tau$ can be extracted from the time dependence of the measured dark count probability in the device. The dark count probability for shorter gate windows can be calculated using $\tau$ and assuming Poisson statistics for the false counts. For a pulse length of 230 ns the dark count probability per pulse is measured as 0.1 at 300K and $10^{-4}$ at $LN_2$. The characteristic time for a dark count, $\tau$ is calculated to be $2.3\times10^{-6}$ (300K) and $2.3\times10^{-3}$ ($LN_2$). Improved dark count probabilities can be achieved with improved processing choices [18, 19].

The probability of success, defined as a Geiger pulse detection being a result of a single arrival and not a dark count, can then be calculated. The logarithmically scaled probability of failure is shown in Fig. 5b. The trade-off between the probability of only one ion strike balanced with the relative probability of a dark count is highlighted by the minima at $\sim 10^{-2}$ ions per gate. This calculation assumes 100% detection efficiency and a false count probability measured for the SIGMA operation at $LN_2$ temperature. A greater than 98% chance of a detected signal corresponding to one and only one ion strike is expected over a range of average number of ions per gate from 0.04 to 0.005. This highlights one significant contribution of this work, which is that a significant improvement in single ion implant confidence can be achieved through reduced temperature SIGMA operation compared to previous room temperature approaches [13].

Improved confidence in single ion implantation can be achieved through reduction of dark count probability. Reducing the gate time, while maintaining the same average ion count per gate, is an immediate way to reduce the probability of dark counts. In this current work we use a gate time of 230 ns, very short duration Geiger mode windows on the order of 1 ns are commonly achieved with active quenching circuitry [14]. Improved device fabrication with diffused junctions, in contrast with the ion implanted junctions in this work, also lead to orders of magnitude smaller dark count probabilities [20, 21]. A further limitation the gating of the ion beam (typically limited to rise and fall time of ~25 ns) could be reduced through the use of novel rapid beam deflection to ~100's of ps. We note that for the present optimum average ions per detection window, an average time of 0.036 seconds would be necessary to implant a single ion with a 98.6% confidence when operating at 2 kHz. Reduced dark count probability would lead to longer average dwell times that might motivate higher detector operation frequencies.

## 4. Conclusion

In summary, we have described single ion detector performance at 300 and 77K that represents a path towards laterally remote, low energy (i.e., minimum straggle), single ion implant detection for single donor device fabrication. Detection of remote single ion strikes at least as far as ~75 μm from the ion detector with 100% detection efficiency is demonstrated. In principle, this lateral sensitivity provides a sufficiently large region to construct future single ion implanted Si devices. The detector sensitivity is bounded, furthermore, at ~600 e-h pairs, which extends ion detector sensitivity to lower energies than previously reported and leads to potentially smaller ion straggle (i.e., improved longitudinal control in atom placement). The ion detector false count probability is significantly decreased below its previously reported 300K performance through the introduction of cryogenic operation. The 77 K operation leads to a drop from $10^{-1}$ to $10^{-4}$ false count probability at 300 and ~77K respectively. This reduction in false count probability has important implications for this approach's ability to reliably implant single ions. The confidence of an implant of one and only one ion arrival is calculated for the dark count probability measured in the devices for this work. An optimal ion flux can be chosen that is predicted to have a 98.6% probability that a detection event corresponds to one and only one ion arrival.

## Acknowledgements

We acknowledge the outstanding assistance from G. Vizkelethy, B. L. Doyle, B. R. McWatters, and K. Childs. This work was supported in full by the National Security Agency Laboratory for Physical Sciences under contract number EAO-09-0000049393. Sandia is a multiprogram laboratory operated by Sandia Corporation, a Lockheed Martin Company, for the United States Department of Energy under Contract No. DE-AC04-94AL85000.